\documentclass[aps,prd,twocolumn,groupedaddress]{revtex4-2}

\usepackage[colorlinks,urlcolor=blue, citecolor=blue,linkcolor=blue]{hyperref}
\usepackage{graphicx}
\usepackage{epsfig}
\usepackage{amsmath}
\usepackage{verbatim}
\usepackage{array}
\usepackage{booktabs}
\usepackage{xcolor}
\usepackage{epstopdf}
\usepackage{color}
\newcommand{\ddstbn}{D^{*0}\bar{D^0}+c.c.}

\begin{document}
\title{Search for the radiative transition $\chi_{c1}(3872)\to\gamma \psi_2(3823)$}

\author{
\begin{small}
\begin{center}
%% Saved at => 2023-10-28
M.~Ablikim$^{1}$, M.~N.~Achasov$^{4,c}$,
P.~Adlarson$^{75}$, O.~Afedulidis$^{3}$, X.~C.~Ai$^{80}$, R.~Aliberti$^{35}$, A.~Amoroso$^{74A,74C}$,
M.~R.~An$^{39}$,
Q.~An$^{71,58,a}$, Y.~Bai$^{57}$
% ...
\\
\vspace{0.2cm}
(BESIII Collaboration)\\
\vspace{0.2cm} {\it
$^{1}$ Institute of High Energy Physics, Beijing 100049, People's Republic of China\\
% ...
$^{3}$ Bochum  Ruhr-University, D-44780 Bochum, Germany\\
$^{4}$ Budker Institute of Nuclear Physics SB RAS (BINP), Novosibirsk 630090, Russia\\
% ...
$^{35}$ Johannes Gutenberg University of Mainz, Johann-Joachim-Becher-Weg 45, D-55099 Mainz, Germany\\
% ...
$^{39}$ Liaoning Normal University, Dalian 116029, People's Republic of China\\
% ...
$^{57}$ Southeast University, Nanjing 211100, People's Republic of China\\
$^{58}$ State Key Laboratory of Particle Detection and Electronics, Beijing 100049, Hefei 230026, People's Republic of China\\
% ...
$^{71}$ University of Science and Technology of China, Hefei 230026, People's Republic of China\\
% ...
$^{74}$ University of Turin and INFN, (A)University of Turin, I-10125, Turin, Italy; (B)University of Eastern Piedmont, I-15121, Alessandria, Italy; (C)INFN, I-10125, Turin, Italy\\
$^{75}$ Uppsala University, Box 516, SE-75120 Uppsala, Sweden\\
% ...
$^{80}$ Zhengzhou University, Zhengzhou 450001, People's Republic of China\\

\vspace{0.2cm}
$^{a}$ Deceased\\
% ...
$^{c}$ Also at the Novosibirsk State University, Novosibirsk, 630090, Russia\\
% ...

}
%% ends here %%

\end{center}
\vspace{0.4cm}
\end{small}
}

\begin{abstract}
Using 9.0 $\rm fb^{-1}$ of $e^+e^-$ collision data collected at center-of-mass 
energies from 4.178 to 4.278 GeV with the BESIII detector 
at the BEPCII collider, we perform the first search for the radiative transition $\chi_{c1}(3872)\to\gamma \psi_2(3823)$.  
No signal is observed and the upper limit on the ratio of branching fractions
$\mathcal{B}(\chi_{c1}(3872)\to\gamma \psi_2(3823), \psi_2(3823)\to\gamma\chi_{c1})/\mathcal{B}(\chi_{c1}(3872)\to\pi^+\pi^- J/\psi)$
is set at 0.075 at the 90\% confidence level. 
Our result contradicts theoretical predictions under the assumption that the $\chi_{c1}(3872)$ is the pure charmonium state $\chi_{c1}(2P)$. 
\end{abstract}.

\maketitle

\section{Introduction}\label{intro}

As the prototypical example of charmonium-like $XYZ$ states, the $\chi_{c1}(3872)$ has been extensively investigated in the past two decades since it was discovered by the Belle experiment~\cite{intx1} in 2003. From a global fit to the measurements by LHCb, BESIII, Belle, BaBar, and others, its mass and width are determined to be $\rm M=3871.65\pm0.06$ MeV/$c^2$ and $\Gamma=1.19\pm0.21$ MeV, respectively~\cite{PDG2022}.  
Its spin, parity and charge-conjugation parity quantum numbers are determined
to be $J^{PC}=1^{++}$~\cite{lhcbq}. So far, the observed decay modes of the particle include 
$\ddstbn$, $\pi^+\pi^- J/\psi$, $\omega J/\psi$, $\gamma J/\psi$, and 
$\pi^0\chi_{c1}$~\cite{Be2011a, Ba2008a, Be2010c, Ba2008c, Ba2010d, Bs1, Bs2, Bs3, Be2011b, Ba2009b}.
Although tremendous effort has been made from both the experimental and theoretical sides,
the interpretation of the $\chi_{c1}(3872)$ remains inconclusive.
Due to the proximity of its mass to the $D^{*0}\bar{D^0}+c.c.$
mass threshold, it is 
conjectured to have a large $D^{*0}\bar{D^0}+c.c.$ molecular component~\cite{intx2,intx22}. Indeed, some theoretical models consider it to be a mixture of 
a conventional $2^3P_1$ charmonium state $\chi_{c1}(2P)$ and a $D^{*0}\bar{D^0}+c.c.$ molecule~\cite{mix_su,mix_zhao}.

Measurements of new $\chi_{c1}(3872)$ decay modes can help to improve our understanding of its internal structure.
Ref.~\cite{Li:2019kpj} extracted the absolute branching 
fractions of the known $\chi_{c1}(3872)$ decays
by performing a global fit of the absolute branching fraction of the $B^+\to\chi_{c1}(3872)K^+$ channel measured by BaBar~\cite{babar_abso} together with information from other experiments. The fraction of $\chi_{c1}(3872)$ decays not observed in experiments is estimated to be $31.9^{+18.1}_{-31.5}$\%.
The work is carried out by assuming the $\chi_{c1}(3872)$
has universal properties in different production and decay mechanisms. 
Meanwhile, Ref.~\cite{Braaten:2019ags} also 
reported the branching fractions with consideration of 
the threshold effect of $\ddstbn$
and a possible bound state below the threshold or a virtual 
state in $B^+\to\chi_{c1}(3872)K^+$ decay.
If the $\chi_{c1}(3872)$ contains a component of 
the excited spin-triplet state $\chi_{c1}(2P)$, 
then the radiative decay $\chi_{c1}(3872)\to\gamma \psi_2(3823)$ 
could happen naturally via a E1 transition~\cite{nr_gi}, where the $\psi_2(3823)$ is considered as the $1^3D_2$ charmonium state. 
The BESIII experiment has reported the observation of $e^+e^-\to\gamma \chi_{c1}(3872)$ at 
center-of-mass energies $\sqrt{s}=4.178-4.278$ GeV~\cite{besx1,besx2}.
Using the $\chi_{c1}(3872)$ signal produced in these data samples, we search for the 
radiative transition $\chi_{c1}(3872)\to\gamma \psi_2(3823)$,
where the $\psi_2(3823)$
is reconstructed with the cascade decay
$\psi_2(3823)\to\gamma\chi_{c1}$,
$\chi_{c1}\to\gamma J/\psi$,
$J/\psi\to \ell^+\ell^-$ ($\ell=e,\mu$).
The branching fraction ratio of this decay relative to the well-established
$\chi_{c1}(3872)\to\pi^+\pi^-J/\psi$ decay, 
$\mathcal{R}_{\chi_{c1}(3872)}\equiv\mathcal{B}(\chi_{c1}(3872)\to\gamma \psi_2(3823), \psi_2(3823)\to\gamma\chi_{c1})/
\mathcal{B}(\chi_{c1}(3872)\to\pi^+\pi^- J/\psi)$, is determined in this work.

Many theoretical models predict the partial widths 
of the radiative transitions between different conventional charmonium states.
The partial widths of $\chi_{c1}(2P)\to\gamma\psi(1^3D_2)$ and 
$\psi(1^3D_2)\to\gamma\chi_{c1}(1P)$
are calculated with the non-relativistic (NR) potential model
and the Godfrey-Isgur (GI) relativistic potential model~\cite{nr_gi}.
Recently, the partial width of $\psi(1^3D_2)\to\gamma\chi_{c1}(1P)$ was
calculated with lattice QCD (LQCD)~\cite{the_lqcd}, and the total width of the $\psi(1^3D_2)$ was estimated according to the BESIII measurements and some phenomenological results. 
Combining these predictions with the total width of the $\chi_{c1}(3872)$, $\Gamma_{\chi_{c1}(3872)}=1.19\pm0.21$ MeV,
we calculated the theoretical branching factions 
$\mathcal{B}(\chi_{c1}(2P)\to\gamma\psi(1^3D_2)$ and $\mathcal{B}(\psi(1^3D_2)\to\gamma\chi_{c1}(1P))$,
and then proceed to the ratio of branching fractions, $\mathcal{R}_{\chi_{c1}(2P)}\equiv\mathcal{B}(\chi_{c1}(2P)\to\gamma\psi(1^3D_2),~\psi(1^3D_2)\to\gamma\chi_{c1}(1P))/\mathcal{B}(\chi_{c1}(3872)\to\pi^+\pi^- J/\psi)$ by taking the branching fraction $\mathcal{B}(\chi_{c1}(3872)\to\pi^+\pi^- J/\psi)=(3.8\pm1.2)\times10^{-2}$
from the PDG~\cite{PDG2022}, as listed in Table~\ref{theo_pre}.
It is worth pointing out that the total width of the $\chi_{c1}(3872)$ measured in experiments is highly dependent on the parameterization of its lineshape. 
The value ($1.19\pm0.21$ MeV) used here is from a global fit to the experimental measurements of the decay mode $\chi_{c1}(3872)\to\pi^+\pi^- J/\psi$ which describe the $\chi_{c1}(3872)$ lineshape with a Breit-Wigner (BW) function. The decay 
$\chi_{c1}(3872)\to D^{*0}\bar{D^0}+c.c.$, however, will distort the lineshape due to the 
proximity of its mass to the $D^{*0}\bar{D^0}+c.c.$ threshold. LHCb 
studied the $\chi_{c1}(3872)$ lineshape with a
Flatt$\rm \acute{e}$ model instead~\cite{lhcb_f}, and determined the full width at half maximum (FWHM) of the lineshape to be $0.22^{+0.07+0.11}_{-0.06-0.13}$ MeV, which 
is much smaller than that obtained from the BW model.
Recently, BESIII performed a coupled-channel analysis of the $\chi_{c1}(3872)$
lineshape and reported a FWHM of $0.44^{+0.13+0.38}_{-0.35-0.25}$ MeV~\cite{bes_f}, consistent with the LHCb result. 
If the FWHM values provided by LHCb and BESIII are used to 
calculate $\mathcal{R}_{\chi_{c1}(2P)}$,
the ratios shown in Table~\ref{theo_pre} will increase significantly.
The experimental measurement of this ratio will help to determine
whether the $\chi_{c1}(3872)$ is the conventional charmonium state, $\chi_{c1}(2P)$.

\begin{table*}
\renewcommand\arraystretch{1.5} 
\caption{
The calculated values for $\mathcal{R}_{\chi_{c1}(2P)}$, by including as input values 
the partial decay widths $\Gamma_{\chi_{c1}(2P)\to\gamma\psi(1^3D_2)}$
and $\Gamma_{\psi(1^3D_2)\to\gamma\chi_{c1}(1P)}$ predicted 
by the NR and GI models and LQCD, 
the total widths, $\Gamma_{\chi_{c1}(3872)}$ and $\Gamma_{\psi_2(3823)}$,
and the branching fraction ${\mathcal{B}(\chi_{c1}(3872)\to\pi^+\pi^- J/\psi)}$.
The $``-"$ means unavailable.
The two values of the ratio for the LQCD case correspond to the results by taking the 
$\Gamma_{\chi_{c1}(2P)\to\gamma\psi(1^3D_2)}$ width from the NR and GI models as input, respectively.
}
\label{theo_pre}
\begin{tabular}{c|p{4.5cm}<{\centering}p{2cm}<{\centering}p{2cm}<{\centering}}
  \hline\hline
  \multicolumn{4}{c}{$\Gamma_{\chi_{c1}(3872)}=1190\pm210$ keV~\cite{PDG2022}}\\
  \multicolumn{4}{c}{$\Gamma_{\psi_{2}(3823)}=520\pm100$ keV~\cite{the_lqcd}}\\
  \multicolumn{4}{c}{$\mathcal{B}(\chi_{c1}(3872)\to\pi^+\pi^-J/\psi)=(3.8\pm1.2)\times10^{-2}$~\cite{PDG2022}}\\
  \hline
  &  NR~\cite{nr_gi}  &GI~\cite{nr_gi} & LQCD~\cite{the_lqcd} \\
 %\cline{2-4}
 %\hline
 $\Gamma_{\chi_{c1}(2P)\to\gamma\psi(1^3D_2)}$ (keV) &  35 &  18 & $-$ \\ 
 $\Gamma_{\psi(1^3D_2)\to\gamma\chi_{c1}(1P)}$ (keV) & 307 & 268 & $337\pm27$\\
\hline
$\mathcal{R}_{\chi_{c1}(2P)}$ & $0.46\pm0.19$ & $0.21\pm0.09$ & $0.50\pm0.21$, $0.26\pm0.11$ \\[1.2ex]
  \hline
  \hline
\end{tabular}
\end{table*}

\section{BESIII Detector and data sets}

The BESIII detector~\cite{BES} has an effective geometrical acceptance of 93\% of $4\pi$.
A helium-based main drift chamber (MDC) immersed in a 1 T solenoidal magnetic field measures 
the momentum of charged particles with a resolution of 0.5\% at 1~GeV/$c$ as well 
as the specific energy loss (d$E$/d$x$) with a resolution better than 6\%.
A CsI(Tl) crystal electromagnetic calorimeter (EMC) is used 
to measure energies and positions of photons, where the energy resolution for a 1.0~GeV photon is about
2.5\% in the barrel and 5.0\% in the end caps.
A plastic scintillator time-of-flight system (TOF), with a time resolution of 80~ps (110~ps) in
the barrel (end cap), is used to identify the particles 
combined with the d$E$/d$x$ information measured in the MDC. In addition, 
a multi-gap resistive plate chamber technology is used in the TOF end cap starting from 2015 to improve the time resolution to 60 ps~\cite{tofend};
the data sets in this work benefit from this improvement except for the
data taken at $\sqrt{s}=4.226$ and 4.258 GeV.  
A muon system interleaved in the steel flux return of the magnet based on resistive plate chambers with 2~cm position
resolution provides powerful information to separate muons from pions.

The $e^+e^-$ collision data collected at $\sqrt{s}=4.178-4.278$ GeV is used in this analysis. The integrated luminosity at each energy point is measured with the Bhabha scattering process with a precision better than 1\%~\cite{lumi} as listed in Table~\ref{DataSet}.
A {\sc geant4}-based~\cite{geant} software package is used to generate
the Monte Carlo (MC) simulated data samples. 
The inclusive MC samples, used to estimate the backgrounds, include the open-charm hadronic processes, continuum processes, and the initial-state-radiation effects, and are generated with {\sc kkmc}~\cite{KKMC} in conjunction with {\sc evtgen}~\cite{evtgen}. 
The signal MC samples $e^{+}e^{-}\to\gamma \chi_{c1}(3872)$, with the decay chain 
$\chi_{c1}(3872)\to\gamma \psi_2(3823)$,
$\psi_2(3823)\to\gamma\chi_{c1}$, $\chi_{c1}\to\gamma J/\psi$, 
$J/\psi\to \ell^+\ell^-$ ($\ell=e,\mu$),
are used to determine the detection efficiency.
The $e^{+}e^{-}\to\gamma \chi_{c1}(3872)$ decay is simulated 
as an E1 transition according to the measurement from BESIII~\cite{besx1}.
The $\chi_{c1}(3872)\to\gamma \psi_2(3823)$ and $\psi_2(3823)\to\gamma\chi_{c1}$ decays are produced with a phase space model.

\begin{table}
\caption{The data sets and their integrated luminosity at each energy point.}
\label{DataSet}
\begin{tabular}{cc}
  \hline\hline
  $\sqrt{s}$ (GeV) &  Luminosity ($\textrm{pb}^{-1}$) \\
  \hline
  4.178& 3189.0\\
  4.189& 526.7\\
  4.199& 526.0\\
  4.209& 517.1\\
  4.219& 514.6\\
  4.226& 1101.0\\
  4.236& 530.3\\
  4.244& 538.1\\
  4.258& 828.4\\
  4.267& 531.1\\
  4.278& 175.7\\
  \hline
  \hline
\end{tabular}
\end{table}

\section{Event Selection and result}

According to the decay chain of the signal process, $e^{+}e^{-}\to\gamma \chi_{c1}(3872)$, $\chi_{c1}(3872)\to\gamma \psi_2(3823)$,
$\psi_2(3823)\to\gamma\chi_{c1}$,
$\chi_{c1}\to\gamma J/\psi$, 
$J/\psi\to \ell^+\ell^-$ ($\ell=e,\mu$),
the final state contains a lepton pair from the $J/\psi$ decay 
and four radiative photons. For the leptons, each corresponding charged track 
is required to have its point of closest approach
to the beam axis within 1~cm in the radial direction and
within 10 cm along the beam direction, and to lie
within the polar-angle coverage of the MDC, $|\cos\theta|<0.93$, in the
laboratory frame.
We require exactly two good charged tracks in the candidate events.
EMC information discriminates between the electrons and muons:
electrons are required to deposit at least 0.8~GeV in the EMC,
and the muons less than 0.4~GeV. 
Photons are reconstructed from isolated showers in the EMC,
at least 10 degrees away from any charged track,
with an energy deposit of at least 25~MeV in both the barrel ($|\cos\theta|<0.80$) and 
the end-cap ($0.86<|\cos\theta|<0.92$) regions. 
In order to suppress electronic noise unrelated to the event, 
the EMC time, $t$, of the photon candidate must be in the range 
$0\leq t\leq700~\rm{ns}$, consistent with collision events.  
We require at least four photons for each candidate event. 

A four-constraint (4C) kinematic fit is applied to constrain the total four-momentum of the lepton pair 
and the four photons to that of the colliding beams, to suppress backgrounds and improve the resolution. 
For events with more than four photons, the combination with the best fit quality corresponding to the minimum fit chi-square, $\chi^2_{\rm 4C}$, is retained. The $J/\psi$ is reconstructed by requiring the invariant mass, $M(\ell\ell)$, 
of the lepton pair to satisfy $|M(\ell\ell)-m(J/\psi)|<30$ MeV/$c^2$,
where $m(J/\psi)$ is the nominal $J/\psi$ mass.
The selection criteria are optimized by maximizing the punzi figure-of-merit 
$S/(\frac{a}{2}+\sqrt{B})$~\cite{pun}, where 
the number of signal events ($S$) 
is determined with the signal MC sample, the background ($B$) 
is estimated with the inclusive MC, and the 
expected statistical significance ($a$) is set to be 3.
The dominant background is from the process $e^+e^-\to\pi^0\pi^0 J/\psi$.
After the $J/\psi$ selection, we veto $\pi^0$ candidates by requiring 
that the invariant mass of all photon pairs is more than 15 MeV/$c^2$ 
away from the nominal $\pi^0$ mass.  
After these requirements, a seven-constraint 
(7C) kinematic fit with an additional three constraints on the masses of 
$M(\ell\ell)$, $M(\gamma\ell\ell)$,
and $M(\gamma\gamma\ell\ell)$ to the nominal masses 
of $J/\psi$, $\chi_{c1}$, and $\psi_2(3823)$, respectively, 
is applied to distinguish the radiative photon 
in each cascade decay.
The best-fit combination with the minimum chi-square, $\chi^2_{\rm 7C}$, 
is retained; $\chi^2_{\rm 7C}<100$ is also required 
to further suppress the combinatorial backgrounds. 
One possible peaking background is $\psi_2(3823)\to\gamma\chi_{c2},~\chi_{c2}\to\gamma J/\psi$, 
the contribution of which is estimated according to the measurement of the branching 
fraction ratio of $\psi_2(3823)\to\gamma\chi_{c2}$ to $\psi_2(3823)\to\gamma\chi_{c1}$
in Ref.~\cite{pc2}. The ratio of the yields of $\psi_2(3823)\to\gamma\chi_{c2}$
to $\psi_2(3823)\to\gamma\chi_{c1,2}$,  
is about 1.5\%, which is taken into account as a source of systematic uncertainty.

\begin{figure}
\includegraphics[scale=0.45]{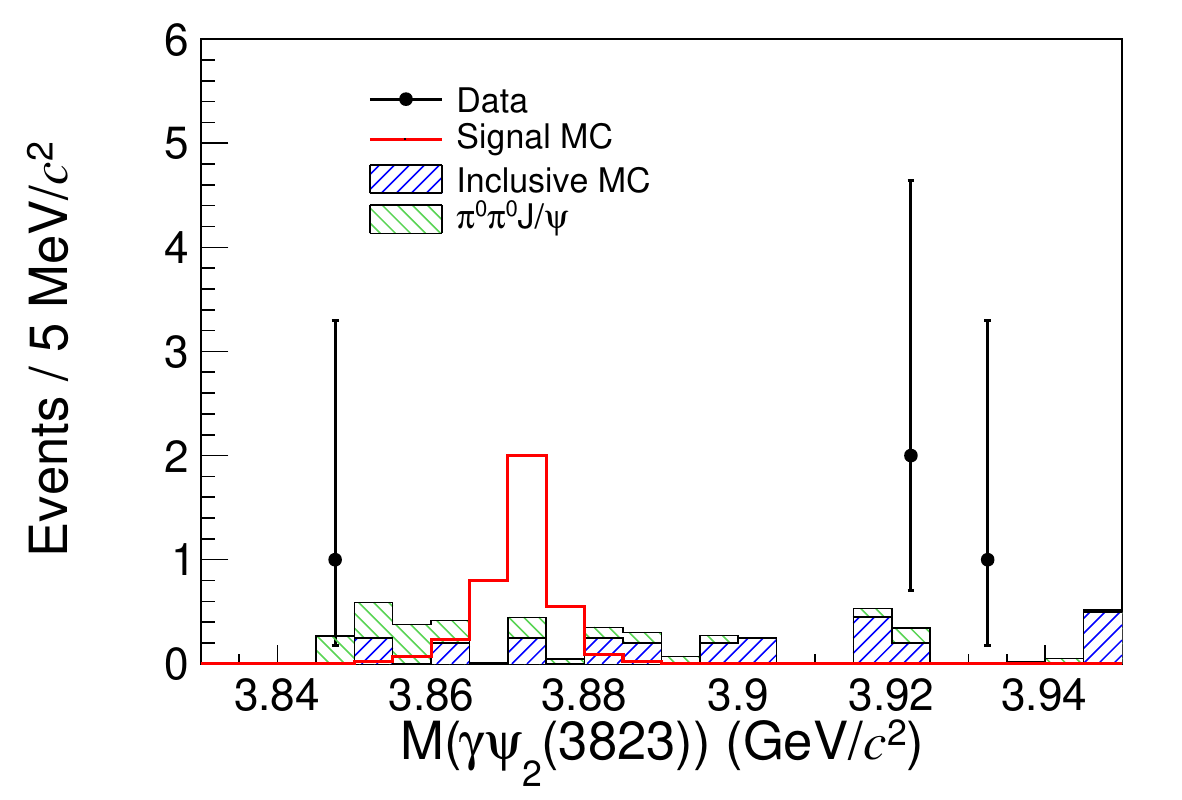}
\caption{Distribution of $M(\gamma \psi_2(3823))$.
The dots with error bars are data, the red 
histogram is the signal MC sample with arbitrary scale, the filled 
blue histogram is the inclusive MC sample without the process $e^+e^-\to\pi^0\pi^0J/\psi$, and
the green stacked histogram is the contribution from $e^+e^-\to\pi^0\pi^0J/\psi$.
}
\label{mx_7c}
\end{figure}

Figure~\ref{mx_7c} shows the distribution of the invariant mass of
the radiative photon and the $\psi_2(3823)$, 
$M(\gamma \psi_2(3823))$ for the selected candidates, 
summed over all the energy points. 
No signal is observed in the $\chi_{c1}(3872)$ signal region in data. 
The three events around 3.93 GeV are very unlikely to be from the 
$\chi_{c2}(2P)$ decays since no $\chi_{c2}(2P)$ signal was observed in its more favourable 
radiative transition to $\psi(2S)$~\cite{Bs1}.
After normalizing the 
MC samples according to the luminosity and cross section in data,
the contributions of the $e^+e^-\to\pi^0\pi^0J/\psi$ process and of the other
backgrounds, estimated with the inclusive MC sample, are also shown in Fig.~\ref{mx_7c}.

\begin{figure}[!h]
\includegraphics[scale=0.45]{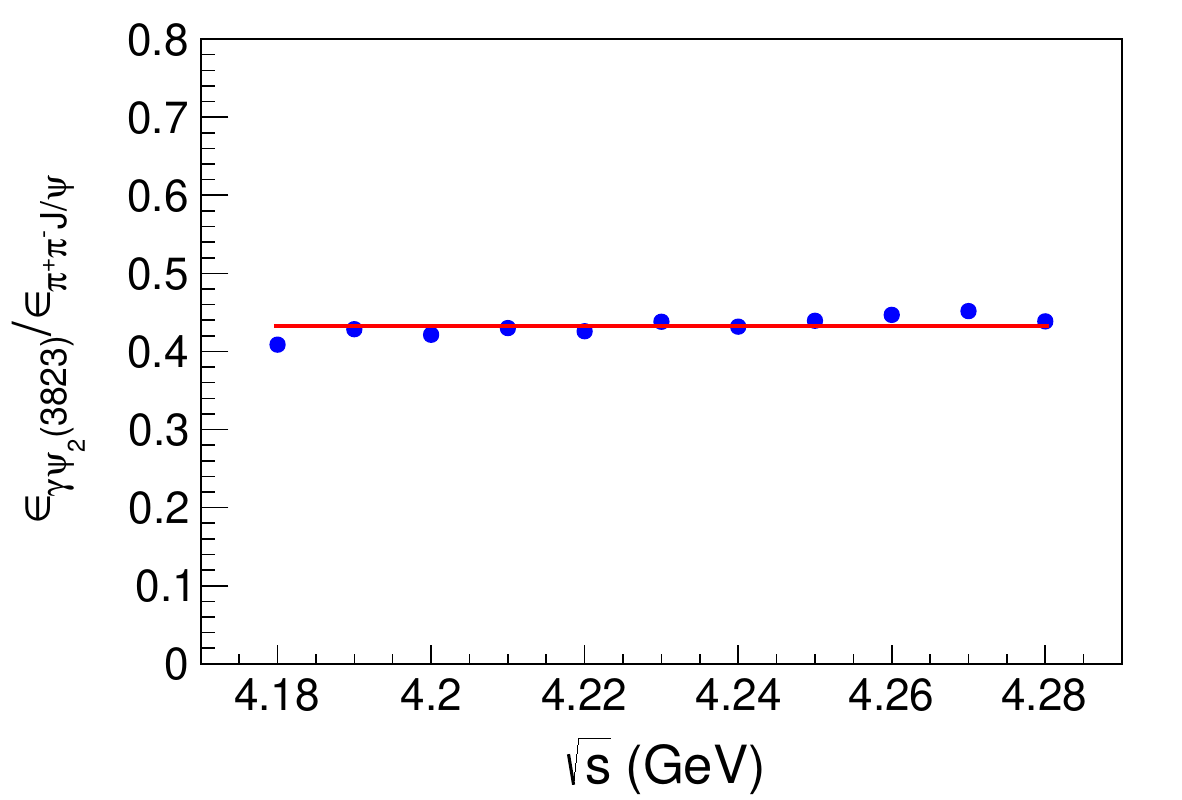}
\caption{
Values of 
$\epsilon_{\gamma \psi_2(3823)}/\epsilon_{\pi^+\pi^- J/\psi}$
at each energy point (blue dots).
The red line indicates the mean value.
}
\label{eff_raa}
\end{figure}

The branching ratio $\mathcal{R}_{\chi_{c1}(3872)}$ is calculated as
\begin{equation}
\begin{aligned}
\mathcal{R}_{\chi_{c1}(3872)}&=\frac{N_{\rm obs}-r\cdot N^{\rm sdb}_{\rm obs}}{N_{\pi^+\pi^- J/\psi}\cdot\frac{\epsilon_{\gamma \psi_2(3823)}}{\epsilon_{\pi^+\pi^- J/\psi}}\cdot\mathcal{B}(\chi_{c1}\to\gamma J/\psi)},
\end{aligned}
\end{equation}
where $N_{\rm obs} = 0$ is the number of observed events
from all data in the $\chi_{c1}(3872)$ signal region
[3.855, 3.885] GeV/$c^2$ which covers around $\pm3\sigma$ of the signal shape according to the signal MC distributions,
$N_{\rm obs}^{\rm sdb} = 4$ is the number of the observed events in the $\chi_{c1}(3872)$ sideband region [3.840, 3.855] and [3.885, 3.940] GeV/$c^2$;
$r$, the background scaling factor from the sideband regions to 
the signal region, is 0.474 based on the inclusive MC sample 
(taking into account its systematic uncertainty; see Sec.~\ref{sec:sys}); 
$N_{\pi^+\pi^- J/\psi}=80.7\pm9.0$ is taken from the BESIII
measurement~\cite{Bs2}; the branching fraction
$\mathcal{B}(\chi_{c1}\to\gamma J/\psi)=0.343\pm0.010$ is quoted from the PDG~\cite{PDG2022};
$\epsilon_{\gamma \psi_2(3823)}$
is the efficiency for the signal process reconstruction, determined with the signal MC sample;
and $\epsilon_{\pi^+\pi^- J/\psi}$ is the efficiency of 
the process $\chi_{c1}(3872)\to\pi^+\pi^- J/\psi$~\cite{Bs2}.
The efficiency ratio $\epsilon_{\gamma \psi_2(3823)}/\epsilon_{\pi^+\pi^- J/\psi}$ at each energy point is shown in Fig.~\ref{eff_raa}, which 
is almost independent on the center-of-mass energy. 
The mean value with the standard deviation,
$\epsilon_{\gamma \psi_2(3823)}/\epsilon_{\pi^+\pi^- J/\psi}=0.433\pm0.004$,
is used to calculate the  $\mathcal{R}_{\chi_{c1}(3872)}$ value. 
The upper limit of  $\mathcal{R}_{\chi_{c1}(3872)}$ at the 90\% confidence level (C.L.) is computed 
with the TRolke program implemented in the ROOT framework~\cite{trolke} by assuming 
the background $N_{\rm obs}^{\rm sdb}$ and the denominator of  $\mathcal{R}_{\chi_{c1}(3872)}$
follow Poisson and Gaussian distributions, respectively, where
the systematic uncertainties discussed in the following section
is taken as the standard deviation of the Gaussian function to be
considered in the upper limit. We obtain an upper limit of
$\mathcal{R}_{\chi_{c1}(3872)}<0.075$ at the 90\% C.L.

\section{systematic uncertainties}\label{sec:sys}

Systematic uncertainties on $\mathcal{R}_{\chi_{c1}(3872)}$ arise mainly from the
estimations of $r$, the possible peaking background of $\psi_2(3823)\to\gamma\chi_{c2}\to\gamma\gamma J/\psi$, $N_{\pi^+\pi^- J/\psi}$,
$\epsilon_{\gamma \psi_2(3823)}/\epsilon_{\pi^+\pi^- J/\psi}$,
and $\mathcal{B}(\chi_{c1}\to\gamma J/\psi)$.
The background scaling factor $r$ is determined from the inclusive MC samples including 
the process $e^+e^-\to\pi^0\pi^0 J/\psi$.
We use a 1st or 2nd order polynomial function to fit the
$M(\gamma\psi_2(3823))$ distribution from the inclusive MC samples;  
the $r$ value is calculated several times using the parameters from the fit
and varying them within $1\sigma$. 
The value $r=0.474$ is chosen from the obtained values 
since it provides the most conservative upper limit.   
The contribution of the potential peaking background of $\psi_2(3823)\to\gamma\chi_{c2}\to\gamma\gamma J/\psi$
is estimated with the related measurements mentioned previously within one errors, 
and the result providing the most 
conservative $\mathcal{R}_{\chi_{c1}(3872)}$ upper limit is retained.

Both statistical and systematic uncertainties of $N_{\pi^+\pi^- J/\psi}$
contribute as sources of systematic uncertainty,
where the statistical part (11.2\%) is obtained
by assuming that $N_{\pi^+\pi^- J/\psi}$ follows a Poisson
distribution, and the systematic part (4.1\%) 
is obtained from Ref.~\cite{Bs2} where the dominant contribution is
from the parametrization of the $\chi_{c1}(3872)$ signal shape. 
The systematic uncertainty (2.9\%) 
due to $\mathcal{B}(\chi_{c1}\to\gamma J/\psi)$ is taken from the PDG~\cite{PDG2022}.
The systematic uncertainty  
of $\epsilon_{\gamma \psi_2(3823)}/\epsilon_{\pi^+\pi^- J/\psi}$
comes mainly from 
the tracking (2.0\%),
the photon selection (3.0\%), and the kinematic fit (2.2\%) uncertainties, estimated 
with the control sample $e^+e^-\to\pi^0\pi^0 J/\psi$.
The systematic uncertainty due to the $\pi^0$ veto is 
mainly caused by potential differences in the angular 
distributions of the radiative photon between the data 
and the signal MC sample, and it is estimated by 
changing the angular distribution of the radiative 
$\gamma$ in $\chi_{c1}(3872)\to\gamma \psi_2(3823)$ 
to $1\pm \cos^2\theta$ (from flat) in the generator model.
The relative difference of 5.3\% between the efficiencies obtained with
the photon angular distributions of $1-\rm cos^2\theta$
and $1+\cos^2\theta$ is taken as the systematic uncertainty.

The systematic uncertainties are listed in Table~\ref{sys_table}.
The total systematic uncertainty is obtained by summing all systematic
uncertainties in quadrature, assuming they are uncorrelated.

\begin{table}
  \caption{The relative systematic uncertainties on  $\mathcal{R}_{\chi_{c1}(3872)}$.
    Systematics on the sideband scaling ratio, $r$, are treated separately
    (see text).}
\label{sys_table}
\begin{tabular}{llr}
  \hline\hline
  Item & &sys. (\%)\\ 
  \hline
$N_{\pi^+\pi^- J/\psi}$  &stat.& 11.2\\
& sys. &4.1\\
\hline
& tracking & 2.0\\
$\epsilon_{\gamma \psi_2(3823)}/\epsilon_{\pi^+\pi^- J/\psi}$ & photon & 3.0\\
& kinematic fit& 2.2\\
& $\pi^0$ veto& 5.3\\
\hline
$\mathcal{B}(\chi_{c1}\to\gamma J/\psi)$& &2.9\\
\hline
Sum&  &14.1\\ 
\hline\hline
\end{tabular}
\end{table}

\section{summary}

In summary, we search for the radiative decay
$\chi_{c1}(3872)\to\gamma \psi_2(3823)$ for the first time by using the 
$e^+e^-$ collision data accumulated at $\sqrt{s}=4.178-4.278$ GeV with the BESIII detector.
No signal is observed, and the upper limit on the branching 
fraction ratio $\mathcal{R}_{\chi_{c1}(3872)}$
is determined to be 0.075 at the 90\% C.L. 
This upper limit is more than $1\sigma$ below 
the theoretical calculations of $\mathcal{R}_{\chi_{c1}(3872)}$ under
the assumption that the $\chi_{c1}(3872)$ is the pure charmonium state $\chi_{c1}(2P)$, listed in Table~\ref{theo_pre},
and much smaller than the predictions based on the FWHMs measured by LHCb and BESIII~\cite{lhcb_f,bes_f}. 
Our result therefore indicates that the $\chi_{c1}(3872)$ is not a pure 
$\chi_{c1}(2P)$ charmonium state.

%\acknowledgments

%% Saved at => 2023-10-28
\textbf{Acknowledgement}

The BESIII Collaboration thanks the staff of BEPCII and the IHEP computing center for their strong support. This work is supported in part by National Key R\&D Program of China under Contracts Nos. 2020YFA0406300, 2020YFA0406400; National Natural Science Foundation of China (NSFC) under Contracts Nos. 11805090, 11635010, 11735014, 11835012, 11935015, 11935016, 11935018, 11961141012, 12025502, 12035009, 12035013, 12061131003, 12192260, 12192261, 12192262, 12192263, 12192264, 12192265, 12221005, 12225509, 12235017; the Chinese Academy of Sciences (CAS) Large-Scale Scientific Facility Program; the CAS Center for Excellence in Particle Physics (CCEPP); Joint Large-Scale Scientific Facility Funds of the NSFC and CAS under Contract No. U1832207; CAS Key Research Program of Frontier Sciences under Contracts Nos. QYZDJ-SSW-SLH003, QYZDJ-SSW-SLH040; 100 Talents Program of CAS; The Institute of Nuclear and Particle Physics (INPAC) and Shanghai Key Laboratory for Particle Physics and Cosmology; European Union's Horizon 2020 research and innovation programme under Marie Sklodowska-Curie grant agreement under Contract No. 894790; German Research Foundation DFG under Contracts Nos. 455635585, Collaborative Research Center CRC 1044, FOR5327, GRK 2149; Istituto Nazionale di Fisica Nucleare, Italy; Ministry of Development of Turkey under Contract No. DPT2006K-120470; National Research Foundation of Korea under Contract No. NRF-2022R1A2C1092335; National Science and Technology fund of Mongolia; National Science Research and Innovation Fund (NSRF) via the Program Management Unit for Human Resources \& Institutional Development, Research and Innovation of Thailand under Contract No. B16F640076; Polish National Science Centre under Contract No. 2019/35/O/ST2/02907; The Swedish Research Council; U. S. Department of Energy under Contract No. DE-FG02-05ER41374.

%% ends here %%

\bibliographystyle{plain}

\end{document}